\documentclass[aps,preprint,tightenlines,nofootinbib,amsmath,amssymb,superscriptaddress]{revtex4}
\usepackage{amssymb}
\usepackage{mathrsfs}
\usepackage{slashed}
\usepackage{graphicx,color}
\usepackage{amsmath}
\allowdisplaybreaks

\begin{document}

\title{TeV Scale Universal Seesaw, Vacuum Stability and Heavy Higgs}
\author{Rabindra N. Mohapatra}
\email{rmohapat@umd.edu}
\affiliation{Maryland Center for Fundamental Physics and Department of Physics, University of Maryland, College Park, Maryland 20742, USA}
\author{Yongchao Zhang}
\email{yongchao@umd.edu}
\affiliation{Center for High Energy Physics, Peking University, Beijing, 100871, P. R. China}
\affiliation{Maryland Center for Fundamental Physics and Department of Physics, University of Maryland, College Park, Maryland 20742, USA}
\date{\today}
\hfill{UMD-PP-014-02}

\begin{abstract}
  We discuss the issue of vacuum stability of standard model by  embedding it within the TeV scale left-right universal seesaw model (called SLRM in the text). This model has only two coupling parameters $(\lambda_1, \lambda_2)$ in the Higgs potential and only two physical neutral Higgs bosons $(h, H)$. We explore the range of  values for $(\lambda_1, \lambda_2)$ for  which the light Higgs boson mass $M_h=126$ GeV and  the vacuum is stable for all values of the Higgs fields. Combining  with the further requirement that the scalar self couplings remain perturbative till typical GUT scales of order $10^{16}$ GeV, we find (i) an upper and lower limit  on the second Higgs $(H)$ mass to be within the range: $0.4 \leq \frac{M_H}{v_R}\leq 0.7$, where the $v_R$ is the parity breaking scale and (ii) that the heavy vector-like top, bottom and $\tau$ partner fermions ($P_3, N_3, E_3$) mass have an upper bound $M_{P_3, N_3, E_3} \leq v_R$. We discuss some phenomenological aspects of the model pertaining to LHC.
\end{abstract}

\maketitle

\tableofcontents

\newpage

\section{Introduction}
The discovery of the 126 GeV Higgs boson at the LHC~\cite{lhc} has raised an interesting issue about the standard model i.e. if there is no new physics below $10^{10}$ GeV or so, the scalar self coupling of the Higgs boson, $\lambda$, when extrapolated by the standard model renormalization group equations, turns negative above this value of mass, primarily due to the large top quark Yukawa coupling. Since mass extrapolation is equivalent to extrapolation to large values of the Higgs field, this will produce a deeper minimum ~\cite{bound} of the Higgs potential compared to the usual vacuum that is used to discuss the physics of the standard model (SM). This leads to instability of the usual vacuum and will lead to the Universe at some time in the future making a transition to the deeper minimum~\cite{transition}. Since the life time for this transition is quite large, one could accept this possibility. However another direction of physics which may be suggested by this and one that has received great deal of attention in recent months, is that this ``near criticality" of the Higgs coupling is a signal of nearby new physics that would stabilize this vacuum. Many extensions of the standard model by enlarging the Higgs and/or the gauge sector have been proposed as a way to eliminate the deeper high scale minimum of the potential~\cite{vacuum}.
\vspace{0.05in}

In this paper, we adopt the second point of view  and consider an extension of the standard model where quark and charged lepton masses arise from a generalized seesaw mechanism (we call it universal seesaw here)~\cite{q-seesaw}, via the introduction of a new set of TeV or higher mass vector-like fermions, that provide the seesaw ``counterweight".  A natural setting for the universal seesaw is not the standard model but one with an extended gauge sector such as the left-right symmetric model based on the gauge group $G_{LR}\equiv SU(2)_L\times SU(2)_R\times U(1)_{B-L}$ with parity symmetry~\cite{LR}. Symmetry breaking in this model is implemented by two Higgs doublets, one, a doublet under $SU(2)_L$ and a second one which is a doublet under $SU(2)_R$. This set up prevents direct Yukawa couplings between the left and right chiral SM quarks that lead to ``non-seesaw" type quark masses, thus enabling seesaw to work. As noted above, the TeV scale vector-like fermion masses provide the seesaw counter weight that leads to masses for the SM quarks as well as charged leptons. There are two options for neutrino masses and we discuss this below. The left-right seesaw model (denoted here by SLRM) has the advantage that it has a particularly simple Higgs sector i.e. only one extra right handed doublet Higgs boson beyond the SM Higgs field. It is therefore different from many multi-Higgs extension of SM discussed in the literature. After symmetry breaking, the model has only two neutral Higgs fields, one of which can be identified with the SM Higgs field ( the 126 GeV Higgs boson). This model has the additional advantage that it also provides a  solution to the strong CP problem without an axion~\cite{babu} and for low right handed scale ($\leq 100$ TeV) , protects~\cite{protect} this solution from possible large Planck scale effects~\cite{planck}. Because of these properties, this model has received a lot of attention in recent years~\cite{ryo}. 

Since the model is based on the gauge group $G_{LR}\equiv SU(2)_L\times SU(2)_R\times U(1)_{B-L}$ with parity symmetry, the Higgs potential of the model has only one extra scalar coupling parameter. The parity symmetry is assumed to be softly broken by the mass terms of the Higgs doublets, so that parity is a technically ``natural" symmetry~\cite{MP}
and does not disturb the above property at the tree level.  As noted, in the unitary gauge, this model has only two neutral Higgs fields and no extra charged Higgs fields. We denote the two Higgs self scalar couplings by $(\lambda_1, \lambda_2)$ and  analyze the renormalization group evolution of these couplings in order to address the stability of the ground state of the theory that breaks the full gauge symmetry down to $U(1)_{em}$. Once a stable vacuum is located, it will solve the vacuum stability problem of SM.

As far as the masses of the heavy vector-like fermions go, in principle, the masses of all but the top partner fermion field could be large but  in this paper for simplicity, we consider both the right handed scale and all the vector-like fermion masse to be in the TeV range in analyzing the vacuum stability of the model. This will make the model amenable to experimental tests at the Large Hadron Collider\footnote{After this work was completed, we came across ref.~\cite{archil}, which discusses vacuum stability in the exact parity symmetric left-right seesaw model with right handed scale as well as the mass of the vector like fermions near $10^{10}$ GeV.  The tree level potential in this model does not break the standard model gauge group, whereas in our case, as we discuss below, parity is softly broken, so that we get a minimum which breaks SM gauge group as well as the right handed $SU(2)_R$ at the tree level. Due to soft parity breaking, left-right symmetry is still ``natural"~\cite{MP}. Also we consider the TeV scale parity breaking as well as vector like fermions with TeV mass so that the second Higgs in our model is observable at the LHC. Our model as well as our discussion of vacuum stability are therefore very different  from the model of~\cite{archil}.}.
\vspace{0.05in}

Main results of our analysis are the following: we find that this simple extension helps to solve the stability problem of SM vacuum and also puts a lower limit on the mass of second neutral Higgs boson of the model. We then impose the requirement that the scalar self couplings do not ``blow up" till the GUT scale of $10^{16}$ GeV, which then allows us to deduce an upper bound  on the second Higgs mass. Combining these we get: $0.4 \leq \frac{M_H}{v_R}\leq 0.7$, where $v_R$ is the parity breaking scale. A second  consequence of vacuum stability requirement is that the heavy vector-like top, bottom and $\tau$ partner fermion ($P_3, N_3, E_3$) masses must have an upper bound, i.e. ${M^{max}_{P_3, N_3, E_3}} \leq v_R$. We then discuss some aspects of the heavy and light Higgs boson phenomenology in the model. We find an interesting relation between the heavy Higgs boson decay modes to $hh, WW, ZZ$ that is characteristic of the model and may be used to test it.
\vspace{0.05in}

This paper is organized as follows: in sec. II, we present the basic ingredients of the model including the scalar potential and the neutral Higgs masses in the unitary gauge; in sec. III, we present the renormalization group equations for different couplings of the model; in sec IV, we present the phenomenology of the heavy Higgs field including its production at LHC and its decay modes. In sec. V, we give some comments on the neutrino mass profiles in our model. We summarize our results in sec. VI.

\section{ Left-right  seesaw model (SLRM)}
\subsection{Particle assignment}
Since this model is a TeV scale embedding of SM in the left-right model with quark and charged lepton seesaw~\cite{q-seesaw}, the SM fermions plus the right handed neutrinos are assigned to doublets of the left and right handed $SU(2)$'s according to their chirality as in standard left-right models. We add four kinds of vector-like fermions $(P, N, E, {\cal N})$, one set per generation, to our model to generate fermion masses\footnote{Strictly speaking, adding the right handed singlet fields ${\cal N}$ is not necessary since the neutrino masses could arise at the loop level~\cite{CM}; in this case, we will get Dirac neutrinos: see later for discussion of this point.}:
\begin{eqnarray}
&& Q_L \in \left( 2,\,1,\,\frac13 \right) \,, \quad
Q_R \in \left( 1,\,2,\,\frac13 \right) \,, \nonumber \\
&& \Psi_L \in \left( 2,\,1,\,-1 \right) \,, \quad
\Psi_R \in \left( 1,\,2,\,-1 \right) \,, \nonumber \\
&& P_{L,\,R} \in \left( 1,\,1,\,\frac43 \right) \,, \nonumber \\
&& N_{L,\,R} \in \left( 1,\,1,\,-\frac23 \right) \,, \nonumber \\
&& E_{L,\,R} \in \left( 1,\,1,\,-2 \right) \,, \nonumber\\
&&{\cal N}_{L,\,R}\in \left(1,\,1,\, 0\right)\,.
\end{eqnarray}
In the above equation, $Q$ and $\Psi$ are the quark and lepton doublets, respectively, and $(Q, P, N)$ are color $SU(3)_c$ triplets while the remaining fields are singlets.
The scalar field content of the left-right seesaw model~\cite{q-seesaw} consists of only of one additional Higgs doublet. They transform under the gauge group $SU(2)_L \times SU(2)_R \times U(1)_{B-L}$ as follows:
\begin{eqnarray}
&& \chi_L =  \left(\begin{matrix}
\chi_L^+ \\ \chi_L^0
\end{matrix}\right)
\in \left( 2,\,1,\,1 \right) \,, \nonumber \\
&& \chi_R =  \left(\begin{matrix}
\chi_R^+ \\ \chi_R^0
\end{matrix}\right)
\in \left( 1,\,2,\,1 \right) \,.
\end{eqnarray}
The scalar  potential in our model is given by:
\begin{eqnarray}
\label{potential}
V &=&
- \mu^2_L \chi_L^\dagger \chi_L
- \mu^2_R \chi_R^\dagger \chi_R \,, \nonumber \\
&& + \lambda_1 \left[ (\chi_L^\dagger \chi_L)^2
    +(\chi_R^\dagger \chi_R)^2 \right]
   + \lambda_2 (\chi_L^\dagger \chi_L) (\chi_R^\dagger \chi_R) \,.
\end{eqnarray}
Note that parity symmetry in the above equation is softly broken so that left-right symmetry is natural~\cite{MP}.
When $\mu^2_{L,R}  > 0$, the full gauge symmetry breaks down to $U(1)_{em}$ at the minimum of the potential:
\begin{eqnarray}
&& \chi_L = \frac{1}{\sqrt2}
\left( \begin{matrix} 0 \\ v_L \end{matrix} \right) \,, \nonumber \\
&& \chi_R = \frac{1}{\sqrt2}
\left( \begin{matrix} 0 \\ v_R \end{matrix} \right) \,,
\end{eqnarray}
we obtain the minimization conditions
\begin{eqnarray}
&& \frac{v_L^2}{2} =
\frac{\lambda_2 \mu_R^2 - 2 \lambda_1 \mu_L^2}{\lambda_2^2 - 4 \lambda_1^2} \,, \nonumber \\
&& \frac{v_R^2}{2} =
\frac{\lambda_2 \mu_L^2 - 2 \lambda_1 \mu_R^2}{\lambda_2^2 - 4 \lambda_1^2} \,.
\end{eqnarray}
Let us note that in ref.\cite{archil},  due to exact left-right symmetry, at the tree level, $v_L=0$ and one has to invoke radiative corrections to generate this vev, which may naturally explain the smallness of $v_L$ compared to $v_R$. This gives additional constraints on the parameters of the model, which we do not have. Also, the tree level soft breaking allows us to keep the right handed scale near a TeV making the right handed $W_R$ accessible at the LHC whereas in ref.~\cite{archil}, the $v_R$ scale is much higher which makes both the $W_R$ and the heavy Higgs $H$ beyond the LHC range. In our model, the smallness of $v_L$ compared to $v_R$ is an input and that allows us to keep the right handed scale in the LHC accessible range.

The CP-even Higgs mass matrix reads
\begin{eqnarray}
\left( \begin{matrix}
2\lambda_1 v_L^2 & \lambda_2 v_L v_R \\
\lambda_2 v_L v_R & 2\lambda_1 v_R^2
\end{matrix} \right) \,.
\end{eqnarray}
In the limit of $v_R \gg v_L$, the two mass eigenvalues are, respectively,
\begin{eqnarray}
\label{Higgs-masses}
&& M_h^2 = 2\lambda_1 \left( 1-\frac{\lambda_2^2}{4\lambda_1^2} \right) v_L^2 \,, \nonumber \\
&& M_H^2 = 2 \lambda_1 v_R^2 \,.
\end{eqnarray}
\subsection{Yukawa interactions and fermion masses}
The Yukawa interactions responsible for fermion masses in this model are given by,
\begin{eqnarray}
- \mathcal{L}_Y &=&
  \bar{Q}_L Y_u \tilde{\chi}_L P_R
+ \bar{Q}_L Y_d \chi_L N_R
+ \bar{\Psi}_L Y_e \chi_L E_R + (L \leftrightarrow R) \nonumber \\
&&+ \bar{P}_L M_P P_R
+ \bar{N}_L M_N N_R
+ \bar{E}_L M_E E_R + {\rm h.c.} \,.
\end{eqnarray}
where $\tilde{\chi}_{L,R} = i\tau_2\chi^*_{L,R}$.
Note that due to the left-right gauge invariance, there is no direct coupling between the left and right handed chiral light quarks as would have been the case for the standard model gauge group with heavy vector-like quarks e.g.~\cite{HY}. We do not include the ${\cal N}$ couplings and discuss it at the end of the paper separately.
In the above equation, $Y_a$ and $M_a$ ($a= u,d,e$) are matrices with complex elements so that theory has CP violation. For simplicity of discussion,  we assume all Yukawa couplings to be all real and note that  our discussion of the Higgs sector and vacuum stability is not affected by this.

\begin{figure}[t]
  \begin{center}
  \includegraphics[width=7.5cm]{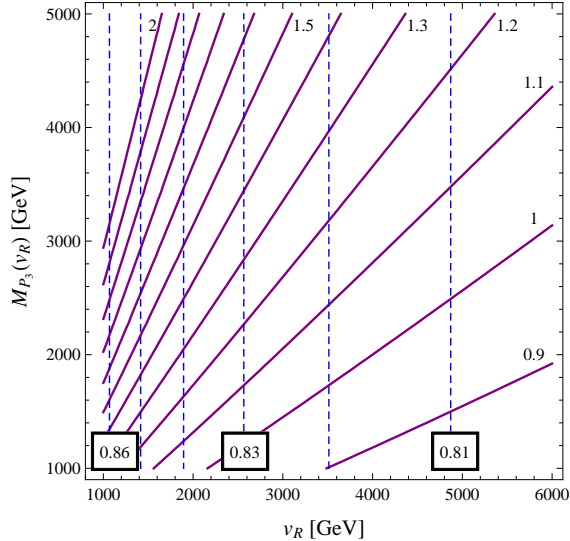}
  \vspace{-.2cm}
  \caption{The purple solid lines indicate the values of $Y_t$ as function of $v_R$ and $M_{P_3}$, and the vertical blue dashed lines are the top quark Yukawa coupling in the SM as function of $v_R$. In making this plot we have used the top quark mass $m_t (m_t) = 163.3 $ GeV~\cite{Xing2}. }
  \vspace{-.7cm}
  \label{Yt}
  \end{center}
\end{figure}

In the SLRM, 
all the quarks obtain their masses from the seesaw mechanism, e.g. for the top sector alone,
\begin{eqnarray}
\left( \begin{matrix}
0 & \frac{1}{\sqrt2} Y_t v_L \\ \frac{1}{\sqrt2} Y_t v_R & M_{P_3}
\end{matrix} \right) \,,
\end{eqnarray}
which leads to generic seesaw type mass relations:
\begin{eqnarray}
\label{seesaw}
m_{q_a}\simeq \frac{Y^2_a v_L v_R}{2M_a} \,.
\end{eqnarray}
Most interesting consequence of the seesaw relation is for the top quark. First of all, the relevant Yukawa coupling $Y_t$ for top quark can differ from that in SM, depending of $v_R$ and the mass of the heavy $P_3$ fermion. For example, if $M_{P_3}\gg v_R$, then $Y_t$ can be much larger than one. In addition to making the theory non-perturbative, large values of $Y_t$ will also lead to gross instability of the vacuum, the very problem we are addressing. We therefore analyze carefully the dependence of $Y_t$ for different values of $v_R$ and $M_{P_3}$. Since we are exploring TeV scale physics, we will keep $v_R$ also in the few TeV range. As shown in Fig.~\ref{Yt}, for $v_R$ and $M_P$ in the range of few TeV, $Y_t$ is generally larger than its corresponding SM value at $v_R$ scale. In combination with RGE analysis, this helps us to put an upper bound on $Y_t$ and hence an upper bound on the top partner mass $M_{P_3}$. We find that in the entire allowed parameter space of our model, $M_{P_3} \leq v_R$.

We now make a brief comment on the quark flavor structure in our model. Let us keep all heavy vector-like fermion masses near the $v_R$ scale. Since the vector-like quark mass matrix can be diagonalized by a choice of basis, all flavor structure resides in the Yukawa coupling matrices $Y_{u,d}$. By taking the Yukawa couplings to be hierarchical, we can reproduce all the quark masses and the CKM angles. We now give an example of a simple scenario which gives the desired CKM mixing pattern. Assume the following: (i) all the mass parameters for the heavy vector-like quarks $P_i$ and $N_i$ are the same with different values for the up and down sector, (ii) the up-type Yukawa coupling matrix $Y_u$ is diagonal with diagonal elements proportional to square roots of $m_{u,c,t}$ and (iii) the down-type Yukawa couplings $Y_d = V_{\rm CKM}^\dagger Y_d^{\rm diag} V_{\rm CKM}$ with $Y_d^{\rm diag}$ proportional to the square roots of $m_{d,s,b}$. This choice naturally leads to the familiar CKM matrix $V_{\rm CKM}$.  One could of course get other fits to observed quark masses and mixings using other choices for vector like quarks with unequal masses etc. The bottom line is that due to extra parameters in the theory beyond SM (e.g. the vector-like quark masses), obtaining desired CKM flavor pattern is always possible in our model.

The presence of the vector-like fermions implies that there are small deviations from unitarity of the CKM matrix in the quark sector; however, the corrections are proportional to ratio of light to heavy quarks and are therefore small. In the simple quark mixing fit, the violation of unitarity is estimated to be at most of order $10^{-4}$. A second point is that
there are new flavor changing neutral current effects in the model due to the presence of TeV scale singlet quarks. These have been analyzed in ref.~\cite{babu}. The resulting constraints on the vector-like quarks are rather weak and imply $M_{P,\,N} \gtrsim$ GeV and can be safely ignored. In the lepton sector, the constraints come from processes such as $\mu\to e+\gamma$ and imply $M_{E} \gtrsim 100$ GeV. The weak limits are due to small ratio of muon to heavy vector-like leptons. The point we want to emphasize is that due to the hierarchical pattern of Yukawa couplings, in the RGE running of couplings, only the third generation fermions make significant contribution. Also we note that CP violating phases do not effect our discussion and therefore we ignore them below.

\section{Renormalization group evolutions (RGE) of couplings and vacuum stability}
In this section, first we present the RGE equations below and above the heavy fermion mass $M_F$ and $SU(2)_R$ symmetry scale $(v_R)$ and then we study their implications for vacuum stability. For simplicity, both the $M_F$ and $v_R$ are chosen to be very near each other and in the TeV range. We use only one matching scale $v_R$ since by virtue of our assumption, all new particles beyond SM start contributing at this scale to the RGEs.

\subsection{RGEs below and above the heavy fermion and right handed scale}

Below the heavy fermion and right handed scale, the SM can be viewed as the effective theory of the SLRM. We therefore use the SM $\beta$ functions till $v_R$ as follows~\cite{RGE:SM}. Note that  our $U(1)_Y$ gauge coupling is not normalized as in GUT theories. \\
\noindent{\it Case I: $\mu \leq v_R, M_F$:}
\begin{eqnarray}
&& \beta(g^\prime) = \frac{1}{16\pi^2}
\left[ \left( \frac{10}{9}n_f +\frac{1}{6} \right)g^{\prime3} \right]  \nonumber\\
&& \beta(g) = \frac{1}{16\pi^2}
\left[ - \left( \frac{43}{6} - \frac23 n_f \right)g^{3} \right] \nonumber\\
&& \beta(g_s) = \frac{1}{16\pi^2}
\left[ - \left( 11 - \frac23 n_f \right)g^{3}_s \right] \nonumber\\
&& \beta(\lambda) = \frac{1}{16\pi^2}
\left[ \frac{9}{8}\left( \frac{1}{3}g^{\prime4} +\frac{2}{3}g^{\prime2}g^{2} +g^{4} \right)
+ 24\lambda^2 -2Y_4
- \lambda (3g^{\prime2} +9g^{2}) +4 \lambda Y_2 \right] \,, \nonumber \\
&& \beta (h_t) = \frac{1}{16\pi^2}
\left[ - h_t \left( \frac{17}{12}g^{\prime2} + \frac94 g^2 + 8g^2_s \right)
+\frac32 h_t (h_t^2-h_b^2)
+ h_t Y_2 \right] \,,\nonumber \\
&& \beta (h_b) = \frac{1}{16\pi^2}
\left[ - h_b \left( \frac{5}{12}g^{\prime2} + \frac94 g^2 + 8g^2_s \right)
+\frac32 h_b (h_b^2-h_t^2)
+ h_b Y_2 \right] \,,\nonumber \\
&& \beta (h_\tau) = \frac{1}{16\pi^2}
\left[ - \frac94 h_\tau \left( \frac53 g^{\prime2} + g^2 \right)
+\frac32 h_\tau^3
+ h_\tau Y_2 \right] \,,
\end{eqnarray}
with $n_f$ the number of flavors, and
\begin{eqnarray}
&& Y_2 = 3h_t^2 + 3h_b^2 + h_\tau^2 \,, \nonumber \\
&& Y_4 = 3h_t^4 + 3h_b^4 + h_\tau^4 \,.
\end{eqnarray}

Above the ($v_R, M_F$) scales (which we assume to be nearly the same), due to the extended gauge interaction and the heavy vector-like fermions, the $\beta$ functions are substantially different (note that we have a different set of Yukawa couplings from the effective SM theory, through they are closely correlated, and see the matching conditions below for the normalization of $g_{BL}$), \\
\noindent{\it Case II:  $\mu \geq v_R, M_F$:}
\begin{eqnarray}
\label{RGE2}
&& \beta(g_{BL}) = \frac{1}{16\pi^2}
\left[ \frac{41}{2} g^{3}_{BL} \right] \nonumber\\
&& \beta(g) = \frac{1}{16\pi^2}
\left[ - \frac{19}{6} g^{3} \right] \nonumber\\
&& \beta(g_s) = \frac{1}{16\pi^2}
\left[ - 3 g^{3}_s \right] \nonumber\\
&& \beta(\lambda_1) = \frac{1}{16\pi^2}
\left[ \frac{9}{8}\left( \frac{3}{4}g^{4}_{BL} +g^{2}_{BL}g^{2} +g^{4} \right)
+ ( 24\lambda^2_1 +2\lambda_2^2 )
-2 \tilde{Y}_4 \nonumber \right. \\
&& \qquad\qquad\qquad\;\;
\left. - \lambda_1 \left(\frac92 g^{2}_{BL} +9g^{2}\right)
+ 4 \lambda_1 \tilde{Y}_2  \right] \,, \nonumber \nonumber\\
&& \beta(\lambda_2) = \frac{1}{16\pi^2}
\left[ \frac{27}{16}g^{4}_{BL}
+ ( 24\lambda_1\lambda_2 +4\lambda_2^2 )
- \lambda_2 \left(\frac92g^{2}_{BL} +9g^{2}\right)
+ 4 \lambda_2 \tilde{Y}_2  \right]  \nonumber\\
&& \beta (Y_t) = \frac{1}{16\pi^2}
\left[ \frac32 Y_t (Y_t^2-Y_b^2)
- Y_t \left( \frac{17}{8}g^{2}_{BL} + \frac94 g^2 + 8g^2_s \right)
+ Y_t \tilde{Y}_2 \right] \nonumber \\
&& \beta (Y_b) = \frac{1}{16\pi^2}
\left[ \frac32 Y_b (Y_b^2-Y_t^2)
- Y_b \left( \frac{5}{8}g^{2}_{BL} + \frac94 g^2 + 8g^2_s \right)
+ Y_b \tilde{Y}_2 \right] \nonumber\\
&& \beta (Y_\tau) = \frac{1}{16\pi^2}
\left[ \frac32 Y_\tau^3
- \frac94 Y_\tau \left( \frac{5}{2} g^{2}_{BL} + g^2 \right)
+ Y_\tau \tilde{Y}_2 \right] \,,
\end{eqnarray}
with
\begin{eqnarray}
&& \tilde{Y}_2 = 3Y_t^2 + 3Y_b^2 + Y_\tau^2 \,, \nonumber \\
&&\tilde{ Y}_4 = 3Y_t^4 + 3Y_b^4 + Y_\tau^4 \,.
\end{eqnarray}

In order to run the couplings to ultrahigh energy scales, we have to match all the seemingly effective SM couplings to that in the full scenario of SLRM. For simplicity and concreteness, the following matchings are considered at the right handed scale $v_R$:
\begin{itemize}
  \item
  Let us start with the gauge couplings. The matching conditions for strong and weak couplings are trivial, while the matching of $U(1)$ gauges is as follows:
  \begin{eqnarray}
  \frac{1}{\alpha_{Y} (v_R)} =
   \frac35 \frac{1}{\alpha_{I_{3R}} (v_R)}
  +\frac25 \frac{1}{\alpha_{BL} (v_R)} \,,
  \end{eqnarray}
  with
  \begin{eqnarray}
  && \alpha_Y = \frac{\tilde{g}^{\prime2}}{4\pi} \,, \nonumber \\
  && \alpha_{I_{3R}} = \frac{g^{2}}{4\pi} \,, \nonumber \\
  && \alpha_{BL} = \frac{\tilde{g}^{2}_{BL}}{4\pi} \,,
  \end{eqnarray}
  where $\tilde{g}'$ and $\tilde{g}_{BL}$ are the normalized couplings in the context of GUT,
  \begin{eqnarray}
  && \tilde{g}' = \sqrt{\frac53} g' \,, \nonumber \\
  && \tilde{g}_{BL} = \sqrt{\frac23} g_{BL} \,.
  \end{eqnarray}
  Below, the normalized $\tilde{g}_{BL}$ is denoted in Eq.~(\ref{RGE2}) simply as $g_{BL}$.

  \item
  To obtain the matching conditions for the quartic scalar couplings $\lambda$ and $\lambda_{1,\,2}$, we integrate out the heavy scalar at the scale of its mass (approximately the right handed scale) from the potential~\cite{archil}. To the linear order of $\frac{v_L}{v_R}$, the mass term, triple coupling term and quartic coupling term point all have the same matching relationship as implied in Eq.~(\ref{Higgs-masses}),
  \begin{eqnarray}
  \label{matching:scalar}
  \lambda (v_R) = \lambda_1 (v_R) \left( 1 -\frac{\lambda_2^2 (v_R)}{4\lambda_1^2 (v_R)} \right) \,.
  \end{eqnarray}
  This simple relation has deeper phenomenological implications than just being superficially the matching condition: it means evidently that, at the right handed scale, $\lambda_1$ is always larger than the SM quartic coupling $\lambda$ (or we can roughly say that $\lambda$ is increased by the SM scalar interacting with its ``right-handed'' partner), which potentially help to solve the stability problem of the SM vacuum.

  \item
  The matching relation for the Yukawa couplings is somewhat straightforward due to the seesaw mechanism Eq.~(\ref{seesaw}),
  \begin{eqnarray}
  \frac{h_f (v_R)}{\sqrt2} \simeq \frac{Y_f^2(v_R) v_R}{2M_F} \,,
  \end{eqnarray}
  with $f = t,\, b,\, \tau$ and $F$ their corresponding heavy partners. In the numerical running of the RGEs, we will resort to the exact relations, as large Yukawa couplings, especially for the top quark, would invalidate such simple approximation.
\end{itemize}

\subsection{Vacuum stability and universal seesaw}
The standard model has only one Higgs field and the stability vacuum requires that the scalar coupling $\lambda$ must satisfy the positivity condition $\lambda(\mu)>0$ for all values of the mass $\mu$ (or equivalently the field since large values of the field are ``equivalent" to large values of $\mu$). However, as is well known when $\lambda$ is extrapolated to large $\mu$ using renormalization group equations, the negative contribution of the top quark coupling turns it negative around $10^{10}$ GeV for $M_h=126$ GeV for which $\lambda(m_W)\simeq  0.131$. This is the vacuum stability problem.

In the SLRM, the presence of the extra ``right handed'' Higgs doublet $\chi_R$ implies a new scalar coupling $\lambda_2$ and the vacuum stability condition requires that not only $\lambda_1 > 0$ but also $2\lambda_1+\lambda_2 > 0$ and both conditions must be maintained for all values of $\mu$ (or Higgs field). As mentioned above, by choosing $\lambda_2$ appropriately, we can increase the value of $\lambda_1$ at  the $v_R$ scale without conflicting with the observed Higgs mass. However, it cannot be made arbitrarily large since it would then hit the Landau pole when extrapolated to the GUT scale. This means that $\lambda_1$ must have an upper bound.

We assume that the lef-right symmetric theory at the TeV scale that we consider here, is a ``low-energy'' effective phenomenological manifestation of some GUT theory at ultrahigh energy scales.  We therefore assume that the couplings remain perturbative only up to generic GUT scale ($10^{16}$ GeV) but not to the higher Planck scale. Note that we do not mean that our model necessarily unifies to a single GUT group at $10^{16}$ GeV. Unification this model is a highly model dependent issue and is beyond the scope of our work. When we get close to the GUT scale, there would appear new fields and new gauge structure that would change the running profile of the couplings depending on the details of the GUT theory.  We do not to explore this detailed GUT completion, since it is beyond the scope of our discussion.

To be specific, in the numerical running, we set the heavy mass parameters for the third generation to be the same, i.e.
\begin{eqnarray}
M_F = M_{P_3} = M_{N_3} = M_{E_3} \,.
\end{eqnarray}
Note that this does not necessarily mean that the three third generation partners have the same mass eigenvalues (especially the mass eigenvalue of the top quark partner is significantly different from the other two), as they also get contribution from mixing with the SM fermions. At $v_R$ scale, with $v_R$ fixed, the Yukawa couplings are solely determined by the value of $M_F$.

\begin{figure}[t]
  \begin{center}
  \includegraphics[width=7.5cm]{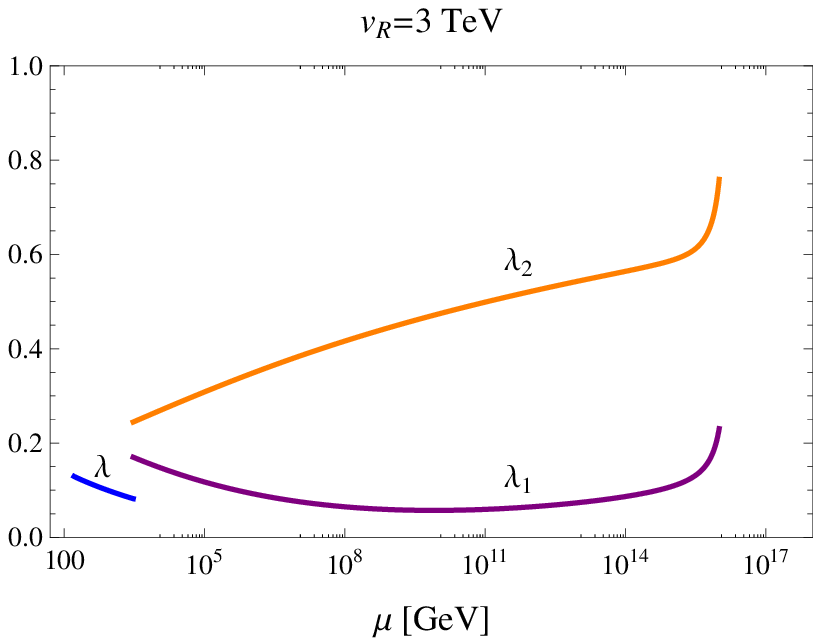}
  \includegraphics[width=7.5cm]{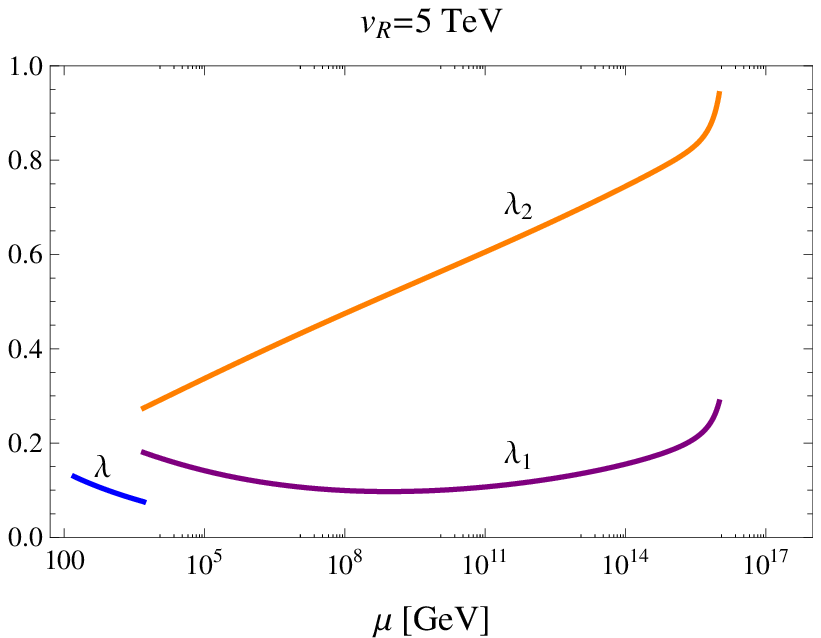}
  \vspace{-.2cm}
  \caption{Examples of running of the quartic couplings $\lambda$ and $\lambda_{1,\,2}$, which are allowed by both the stability and perturbativity constraints. Left: we set $v_R = 3$ TeV, $\lambda_1(v_R) = 0.17$, and $M_F = 1.2$ TeV. Right: we set $v_R = 5$ TeV, $\lambda_1(v_R) = 0.18$, and $M_F = 2$ TeV. For simplicity, we assume $M_F/v_R$ to be nearly same for simplicity in both figures. We have of course chosen the Yukawa coupling parameters in accord with this choice.}
  \vspace{-.7cm}
  \label{running}
  \end{center}
\end{figure}

\begin{figure}[t]
  \begin{center}
  \includegraphics[width=7.5cm]{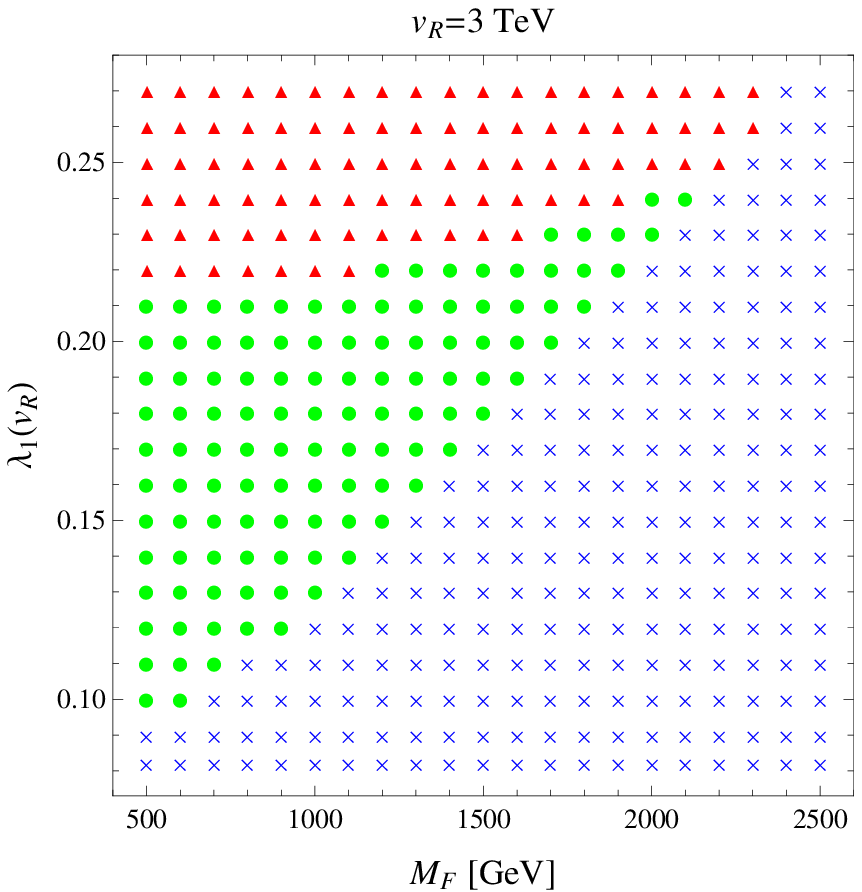}
  \includegraphics[width=7.5cm]{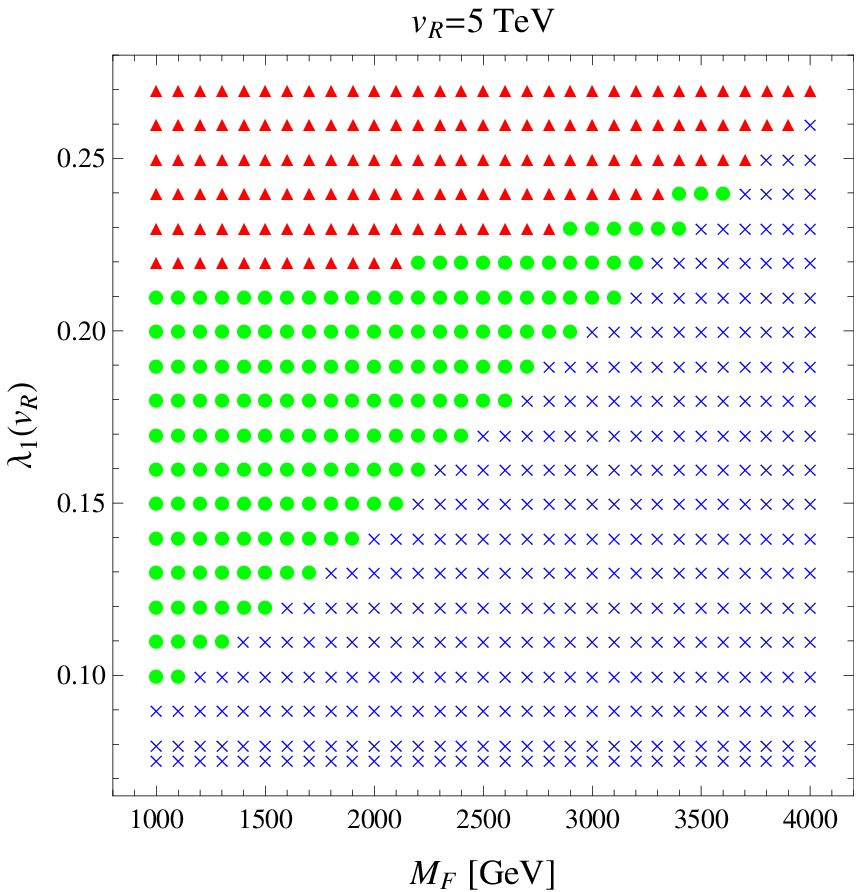}
  \vspace{-.2cm}
  \caption{Scanning the parameter space of SLRM with $v_R = 3$ TeV (left) and 5 TeV (right). The green circles denote the allowed points in the parameter space, blue crosses are excluded by vacuum stability and red triangles excluded by the requirement of perturbativity.}
  \vspace{-.7cm}
  \label{scan}
  \end{center}
\end{figure}

Given a value of $v_R$, we have only two free parameters in the SLRM: the quartic coupling $\lambda_1$ ($\lambda_2$ is fixed by the SM Higgs mass), and the universal heavy fermion mass parameter $M_F$. We also assume the masses of the other generation vector-like fermion masses to be the same as the third generation one but their Yukawa couplings are small and therefore they do not affect our results. We scanned the full parameter space, varying $v_R$ (near the TeV scale), $\lambda_1(v_R)$ and $M_F$. Two examples with completely stable vacuum and perturbative couplings are shown in Fig.~\ref{running} and the scanning results with $v_R = 3$ TeV and 5 TeV are presented in Fig.~\ref{scan}.
Details of the calculation procedure are as follows:
\begin{itemize}
  \item
  We first run the SM couplings from the running top quark mass $m_t (m_t) = 163.3$ GeV~\cite{Xing2} to the right-handed scale $v_R$ and then from the $v_R$ scale to the GUT scale.
  \item
  Values of the electromagnetic and strong gauge couplings, the weak mixing angle and the third generation Yukawa couplings are taken from~\cite{Xing1,Xing2}. We set the SM vev $v = 246$ GeV and use the SM Higgs mass being 126 GeV.
  \item
  We examine the stability of the vacuum for the whole range of running imposing the requirement that $\lambda_1 > 0$ \& $2\lambda_1 + \lambda_2 >0$. We have also checked step by step in the running whether the validity of perturbation theory is respected: $\lambda_1 < 3$ (any large value of other couplings would definitely lead to a large $\lambda_1$). With the energy scale $\mu$ raised up to the ultrahigh value, once either of the conditions is violated, the corresponding point in the parameter space is abandoned.
\end{itemize}

\begin{figure}[t]
  \begin{center}
  \includegraphics[width=7.5cm]{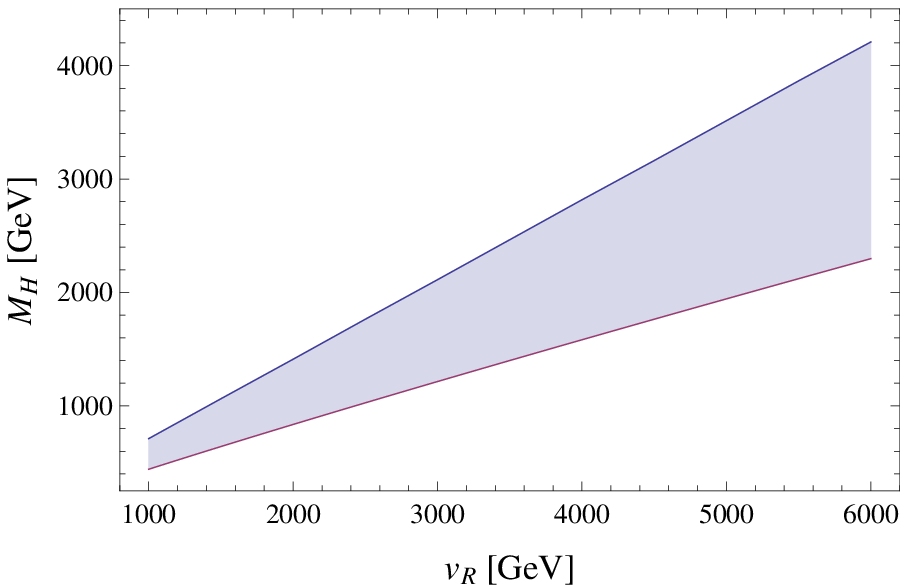}
  \includegraphics[width=7.5cm]{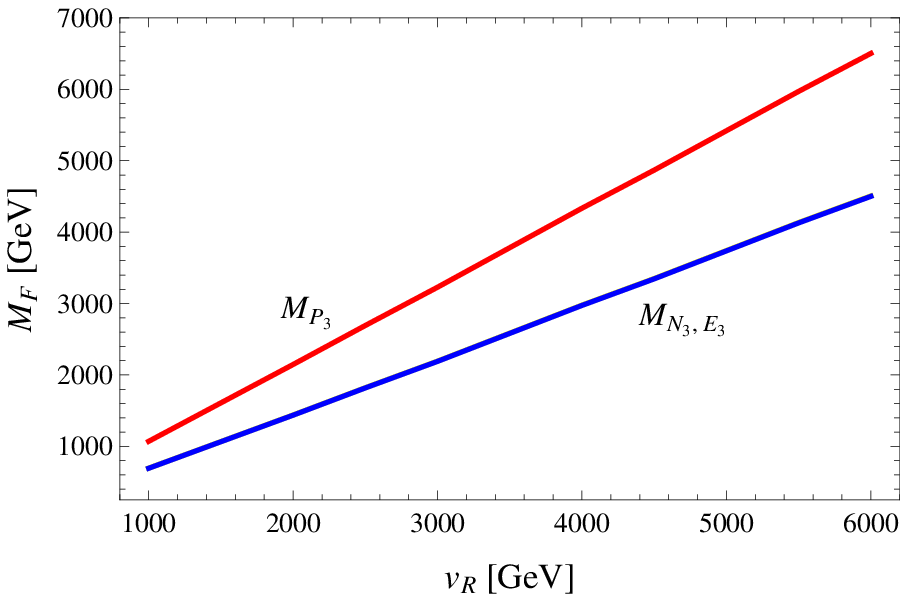}
  \vspace{-.2cm}
  \caption{Left: Constraints on the heavy Higgs mass $M_H$ as function of $v_R$ (the shaded region is allowed) from vacuum stability and perturbativity. Right: Upper bounds on the masses $M_{P_3}$ and $M_{N_3,\,E_3}$ of heavy vector-like fermions as function of $v_R$.}
  \vspace{-.7cm}
  \label{MH+MF}
  \end{center}
\end{figure}

Scanning of the full parameter space reveals first that at $v_R$ scale, the quartic coupling $\lambda_1$ is severely constrained: $\lambda(v_R) < \lambda_1(v_R) \lesssim 0.25$. As pointed out above and shown in Fig.~\ref{scan}, the value of $\lambda_1$ has to be large enough to compensate the negative contributions of $Y_t$ to $\beta(\lambda_1)$, and yet small enough to keep out of the non-perturbative region. This constraint implies that the heavy Higgs mass is predicted to be in the range of about $[\sqrt{2\times0.1},\, \sqrt{2\times0.25}]\, v_R \simeq [0.4, \, 0.7]\, v_R$. We also find that the upper limit on this ratio is nearly independent of $v_R$, while the lower limit has a weak dependence on $v_R$ and $M_F$ (for smaller $v_R$ the lower limit is increased somewhat). The two plots in Fig.~\ref{scan} tell us further that when the masses of heavy vector-like fermion are taken into consideration (e.g. these masses are also constrained), the constraint on the heavy Higgs mass is more stringent. All these facts point to the phenomenological implication that there exists a heavy Higgs in the SLRM at the TeV scale, as explicitly depicted in Fig.~\ref{MH+MF}. In the plot we considered only the constraints from vacuum stability and perturbativity, but not that from the heavy fermion masses. It is interesting that the heavy Higgs boson in the SLRM is potentially detectable at the LHC (and in future high energy colliders); in the next section we will study the LHC phenomenology of this predicted new particle.

\begin{figure}[t]
  \begin{center}
  \includegraphics[width=7.5cm]{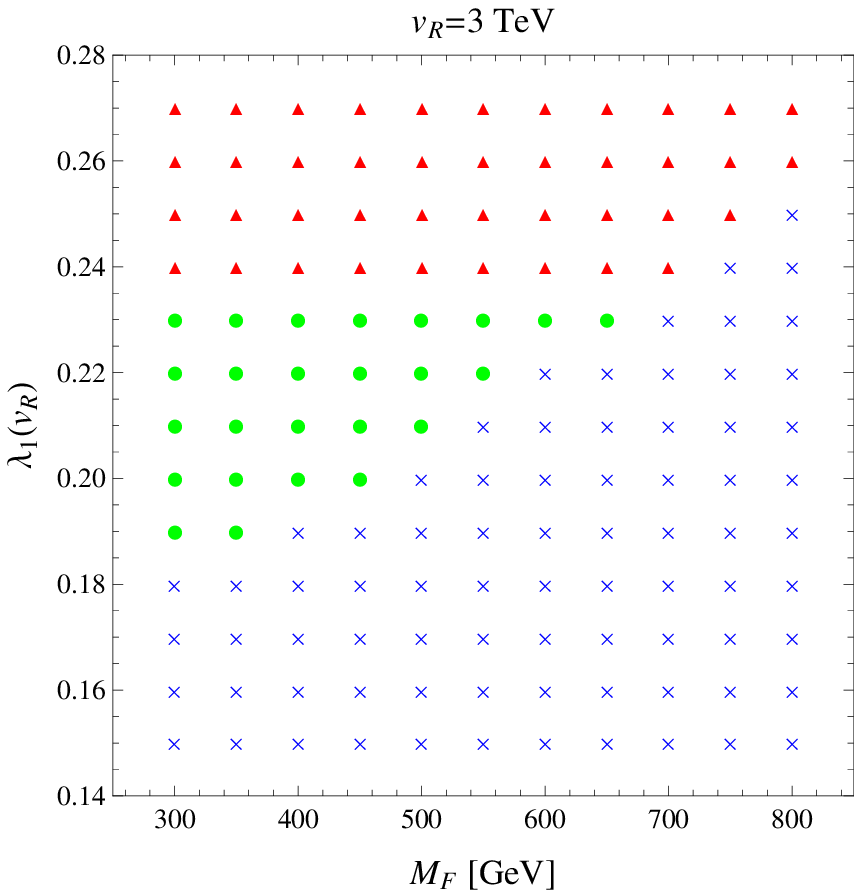}
  \includegraphics[width=7.5cm]{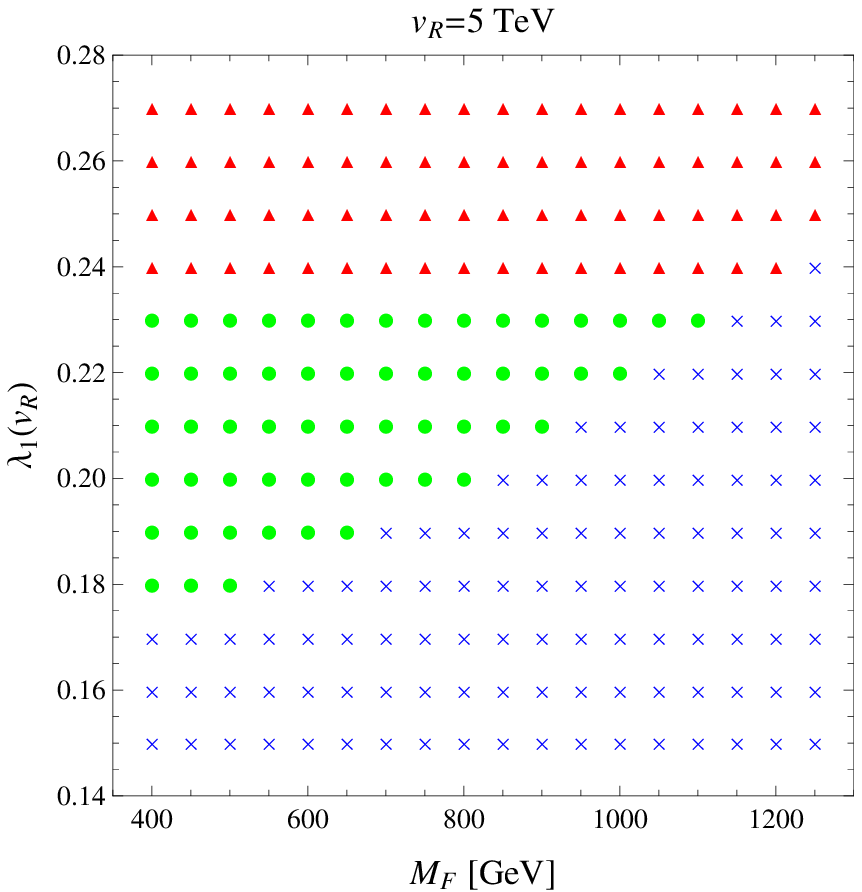}
  \vspace{-.2cm}
  \caption{Scanning the parameter space of SLRM with $v_R = 3$ TeV (left) and 5 TeV (right) assuming $\lambda_2(v_R) < 0$. The markers have the same meaning as in Fig.~\ref{scan}.}
  \vspace{-.7cm}
  \label{scan-}
  \end{center}
\end{figure}

We stress here that the constraints given above are obtained with a positive $\lambda_2$ from Eq.~(\ref{matching:scalar}). We also examined the case with a negative $\lambda_2$ since Higgs mass does not depend on the sign of $\lambda_2$. As expected, negative $\lambda_2$ tends to push the vacuum towards instability, worsening the SM stability problem. Thus the allowed parameter space shrinks greatly. To keep the stability conditions up to the GUT scale, $M_F$ is required to be small. As the examples given in Fig.~\ref{scan-} show, if $v_R = 3$ TeV it is required $M_F \lesssim 650$ GeV while for $v_R = 5$ TeV we get $M_F \lesssim 1100$ GeV. The ATLAS and CMS collaborations have searched for vector-like quarks both with charges $2/3$ and $-1/3$~\cite{Tatlas,Tcms,Batlas,Bcms}, and the most stringent bound at the moment on our model is $M_B \gtrsim 590$ GeV ($B$ is the vector-like quark with charge $-1/3$)~\cite{Batlas}, which sets a lower limit on the negative $\lambda_2$ case: $v_R \gtrsim 2.8$ TeV. With the future search for vector-like quarks at 14 TeV LHC~\cite{Tcms14}, the limit could get much stronger. Comparatively, the positive case is much less constrained and thus phenomenologically preferred. Thus we consider mainly the positive case in this work.

It is transparent in Fig.~\ref{scan} that the heavy fermion mass parameter $M_F$ is also bounded from above, since increasing $M_F$ requires larger $Y_t$ which worsens the stability problem when combined with blow-up far below the Planck scale (upper limit on $\lambda_1$). It has important phenomenological significance since it predicts $M_F < v_R$, or the existence of heavy fermions, the heavy partners of $b$ and $\tau$ fermion, below the right handed scale. This is presented in the right panel of Fig.~\ref{MH+MF}. This coincides with the findings of Ref.~\cite{archil} although these are strictly two different scenarios within the left-right framework. The fact that the top quark partner mass is significantly larger than the other two partners is due to the large top quark Yukawa coupling, which contribute substantially to the top partner mass. In contrast, the contribution to the masses of $N_3$ and $E_3$ from mixing with the SM partners are much smaller and can be safely neglected.

\section{Light and heavy Higgs phenomenology}
In this section, we discuss the implications of the model for heavy (H) and light 126 GeV Higgs boson ($h$) for collider phenomenology.

\subsection{$h$-decay}

As mentioned in the previous section, below the right-handed scale, all the new heavy particles beyond SM (the gauge bosons, the heavy Higgs and the vector-like fermions) are integrated out, and the SM is left as the low energy effective theory. The effects of new physics on SM Higgs decay can be generally neglected, at least to the next-to-leading order of $v_L/v_R$, e.g. for the bottom quark channel,
\begin{eqnarray}
  -\mathcal{L} &\simeq&
    \frac{1}{\sqrt2} \bar{b}_L Y_b h B_R
  + \frac{1}{\sqrt2} \bar{B}_L Y_b H b_R +{\rm h.c.} \nonumber \\
  &\Rightarrow&
  \frac{1}{\sqrt2} \sin\alpha_R^b \, \bar{b}_L^m Y_b h b_R^m +{\rm h.c.} \,.
\end{eqnarray}
Here $B=N_3$ is the heavy partner, $b^m$ the bottom mass eigenstate and $\alpha_R^b$ the right mixing angle of bottom quark with its heavy partner. Approximately $\sin\alpha_R^b \simeq \frac{1}{\sqrt2} Y_b v_R / M_F$ and we recover the SM bottom quark Yukawa coupling via the seesaw relation $\frac{1}{\sqrt2} y_b = Y_b^2 v_R / 2M_F$. For the top quark coupling, although the seesaw relation might not be a good approximation (for $Y_t v_R \sim M_F$), a more exact formula reveals that we can obtain again the same Yukawa coupling as in SM. Phenomenologically, the gluon fusion production and di-photon production processes, in which the top quark loop plays an important role, are not affected in the SLRM\footnote{The heavy top partner loop is suppressed by the small scale mixing angle or left-handed fermion mixing angle (the angle $\alpha_L^t$ used below) and its contribution can be neglected without any effect.}.

\subsection{Triple Higgs coupling}
Another possible effect of beyond the standard model physics is on the triple Higgs coupling~\cite{nir}. To see if there is any such effect, let us
 define the unitary mixing matrix that diagonalizes the mass matrix of the two Higgs bosons as
\begin{eqnarray}
\left( \begin{matrix} h \\ H \end{matrix} \right) = U
\left( \begin{matrix} h_L \\ h_R \end{matrix} \right) \,.
\end{eqnarray}
The equation giving $U$ is
\begin{eqnarray}
U \cong
\left( \begin{matrix} 1 & -\frac{\lambda_2}{2\lambda_1} \frac{v_L}{v_R} \\ \frac{\lambda_2}{2\lambda_1} \frac{v_L}{v_R} & 1 \end{matrix} \right) \,.
\end{eqnarray}
From this, we get the triple couplings from the potential
\begin{eqnarray}
&& \lambda_1 \left[ v_L h_L^3 + v_R h_R^3 \right]
+ \frac12\lambda_2 \left[ v_L h_L h_R^2 + v_R h_L^2 h_R \right] \nonumber \\
& \Rightarrow &
\lambda_1 v_L h^3  \left[ 1- \left( \frac{\lambda_2}{2\lambda_1} \right)^2 \right] \,.
\end{eqnarray}
With the relation given in Eq.~(\ref{Higgs-masses}), the triple coupling is the same as in SM.

\subsection{Production and decay of the Heavy Higgs at LHC}

The decay channels of the heavy Higgs in the SLRM model are given below. We discuss them one by one.
\begin{itemize}
  \item
  $H \rightarrow hh$: for the scalar channel, the LO coupling $m_{Hhh} Hhh$ is given by, $m_{Hhh} \simeq \frac12 \lambda_2 v_R$,
  with the exact value
  \begin{eqnarray}
  m_{Hhh} =
  \frac{1}{2} \varepsilon  \Big( 6\lambda_1 + \left(\varepsilon^2-2\right) \lambda_2  \Big) v_L
  +\frac12 \Big(6 \varepsilon^2 \lambda_1 +\left(1 -2\varepsilon^2 \right) \lambda_2\Big) v_R \,,
  \end{eqnarray}
  where $\varepsilon = \frac{\lambda_2}{2\lambda_1} \frac{v_L}{v_R}$ is the mixing of ``left-handed'' and ``right-handed'' scalars.
  The decay width is then given by
  \begin{eqnarray}
  \Gamma (H \rightarrow hh) =
  \frac{1}{8\pi} \frac{m_{H hh}^2}{M_{H}}
  \left( 1 - \frac{4m_h^2}{M_{H}^2} \right)^{1/2} \,.
  \end{eqnarray}

  \item
  $H \rightarrow t\bar{t}$: for the fermion channel, we assume that the heavy Higgs boson is not heavy enough to decay to the vector-like fermion pairs but decays only into the SM fermions (this corresponds to a large region in the parameter space and there is no fine-tuning for the assumption). Amongst the couplings to the SM fermions, the top quark is expected to be the largest one. We start with the original Lagrangian given below:
  \begin{eqnarray}
  \label{topyukawa}
  -\mathcal{L} &=&
    \frac{1}{\sqrt2} \bar{t}_L Y_t h_L T_R
  + \frac{1}{\sqrt2} \bar{T}_L Y_t h_R t_R +{\rm h.c.} \nonumber \\
  &\Rightarrow&
  \frac{1}{\sqrt2} \bar{t}_L^m H t_R^m \cdot Y_t
  \Big( \varepsilon \cos\alpha_L^t \sin\alpha_R^t
  + \sin\alpha_L^t \cos\alpha_R^t \Big) \nonumber \\
  &\simeq&
  \frac{1}{\sqrt2} \bar{t}_L^m H t_R^m \cdot Y_t
  \Big( \varepsilon \sin\alpha_R^t
  + \sin\alpha_L^t \cos\alpha_R^t \Big) \,,
  \end{eqnarray}
  Here $T\equiv P_3$ is the top quark partner and $t^m$, the mass eigenstate. For a large top Yukawa coupling, the left-handed mixing $\alpha_L^t$ is generally very small, but the right-handed one $\alpha_R^t$ is always very large (generally of order one), since $Y_t v_R \sim M_F$. Denoting the Yukawa coupling $y_{Ht\bar{t}} = Y_t
  ( \varepsilon \sin\alpha_R^t + \sin\alpha_L^t \cos\alpha_R^t )$ which is suppressed by the scalar mixing $\varepsilon$ or left-handed mixing $\alpha_L^t$, the decay width is given by
  \begin{eqnarray}
  \Gamma (H \rightarrow t\bar{t}) =
  \frac{3}{16\pi} \cdot y_{H t\bar{t}}^2 {M_{H}}
  \left( 1 - \frac{4m_t^2}{M_{H}^2} \right)^{3/2} \,.
  \end{eqnarray}

  \item
  $H \rightarrow WW,\,ZZ$: In the SLRM, the gauge bosons $W_L$ and $W_R$ do not mix at tree level, but the scalars do; thus we can get the suppressed coupling $m_{HWW} = 2 \varepsilon M_W^2/v$ with $v = v_L$ being the SM electroweak scale. For the decay width, we get
  \begin{eqnarray}
  \Gamma (H \rightarrow WW) =
  \frac{1}{8\pi} \frac{m_{HWW}^2}{M_{H}}
  \left( 1 - \frac{4m_W^2}{M_{H}^2} \right)^{1/2}
  \left[ 1 + \frac12 \left( 1 - \frac{M_{H}^2}{2m^2_W} \right)^2 \right] \,.
  \end{eqnarray}
  The width for the $ZZ$ boson channel is similar (through the neutral gauge bosons $Z$ and $Z'$ mix at tree level but the mixing is suppressed by $(v_L/v_R)^2$),
  \begin{eqnarray}
  \Gamma (H \rightarrow ZZ) =
  \frac{1}{16\pi} \frac{m_{HZZ}^2}{M_{H}}
  \left( 1 - \frac{4m_Z^2}{M_{H}^2} \right)^{1/2}
  \left[ 1 + \frac12 \left( 1 - \frac{M_{H}^2}{2m^2_Z} \right)^2 \right] \,,
  \end{eqnarray}
  with $m_{HZZ} = 2 \varepsilon M_Z^2/v$.
\end{itemize}

\begin{figure}[t]
  \begin{center}
  \includegraphics[width=6.5cm]{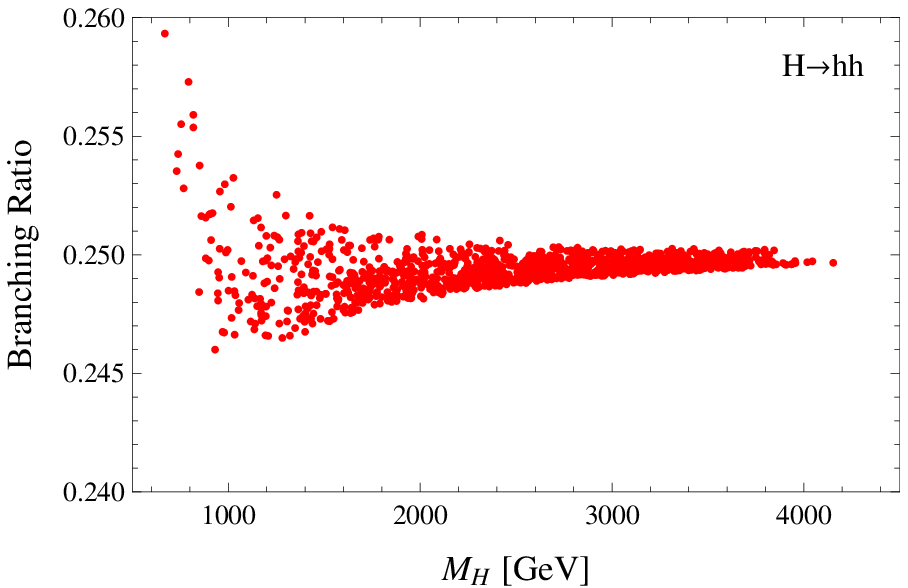}
  \includegraphics[width=6.5cm]{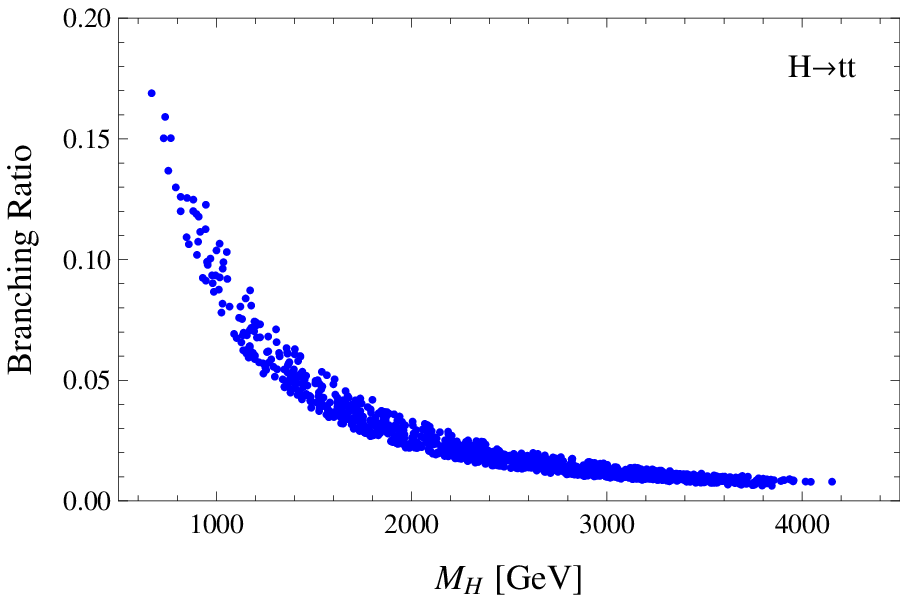}
  \includegraphics[width=6.5cm]{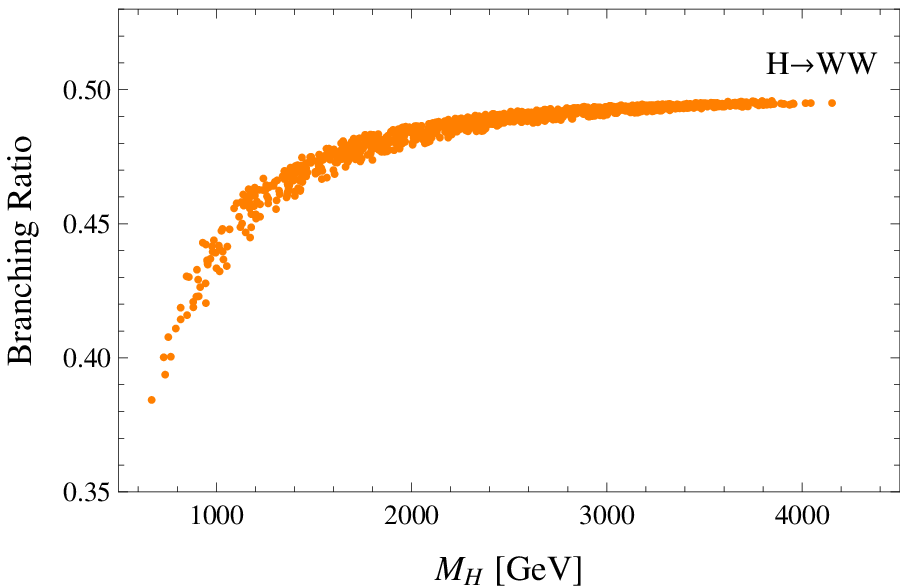}
  \includegraphics[width=6.5cm]{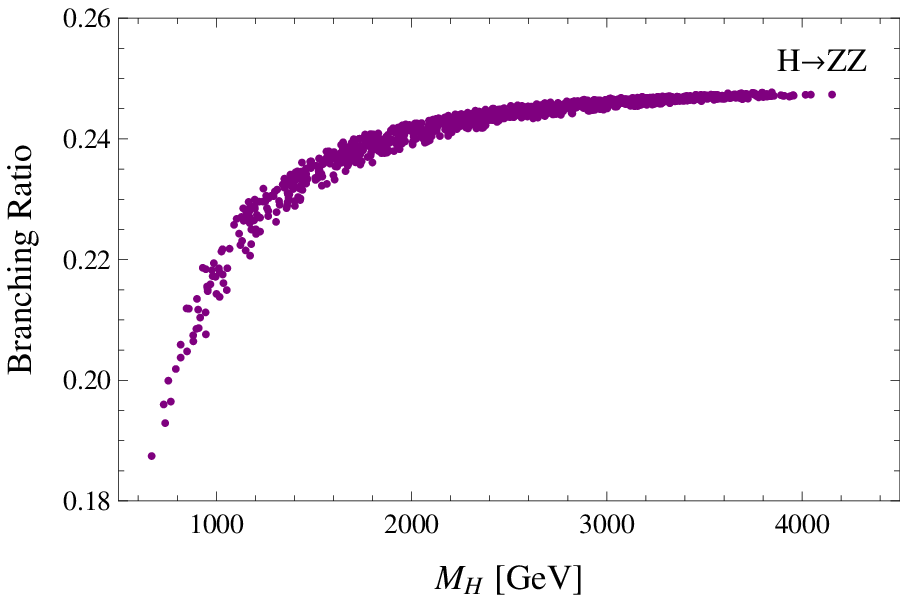}
  \vspace{-.3cm}
  \caption{Branching ratios of heavy Higgs decay. In these plots we do not include the cases in which the heavy fermion channel pair channel(s) is kinematically allowed.}
  \vspace{-.7cm}
  \label{Hdecay}
  \end{center}
\end{figure}

As the heavy Higgs boson is expected to be close to the right-handed scale, which is much larger than the electroweak scale, we can approximate the decay widths and see what happens in the massive limit $v_R \rightarrow \infty$. In this limit, the fermion channel is suppressed by $(v_L/v_R)^2$ as long as $M_F \sim v_R$, while the expression for other channels are very simple, determined only by the parameters $v_R$, $\lambda_1$ and $\lambda_2$,
\begin{eqnarray}
&& \Gamma (H \rightarrow hh) = \frac{1}{8\pi}
\frac{\lambda_2^2}{4 \sqrt{2\lambda_1}} \, v_R \,, \nonumber \\
&& \Gamma (H \rightarrow WW) = \frac{1}{8\pi}
\frac{\lambda_2^2}{2 \sqrt{2\lambda_1}} \, v_R \,, \nonumber \\
&& \Gamma (H \rightarrow ZZ) = \frac{1}{8\pi}
\frac{\lambda_2^2}{4 \sqrt{2\lambda_1}} \, v_R \,.
\end{eqnarray}
The suppression factor $\varepsilon$ for the gauge boson channels is  cancelled by the large enhancement factor $M_H^4/M^4_{W(Z)}$ from interaction with the longitudinal components of gauge bosons. Ultimately it is from the scalar interaction and is therefore not suppressed as these Goldstone bosons are ``eaten" by the gauge bosons. In this limit, we find a relation among these different decay widths, which we call ``the quartering rule'' of heavy Higgs decay, whose origin lies in the coupling of $H$ with the four components of $\chi_L$ before electroweak symmetry breaking. This is explicitly presented in Fig.~\ref{Hdecay}. This extraordinary feature could be a smoking gun signal of the SLRM.
The di-photon channel of the Heavy Higgs decay $H \rightarrow \gamma\gamma$ is predominately mediated by the right-handed $W$ boson, the top quark and its heavy partner. Numerical calculation reveals that the branching ratio of this channel is generally of order $10^{-5}$. Even if the heavy Higgs is observed at colliders, it will be challenging to detect in this specific channel.

\begin{figure}[t]
  \begin{center}
  \includegraphics[width=7.5cm]{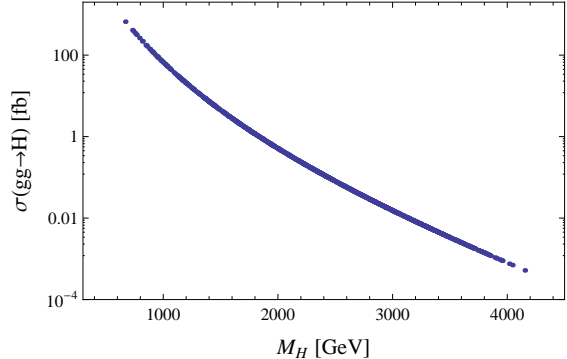}
  \vspace{-.3cm}
  \caption{Heavy Higgs production cross section $\sigma(gg \rightarrow H)$ at LHC with center-of-mass energy of 14 TeV, as function of $H$ mass.}
  \vspace{-.7cm}
  \label{Hproduction}
  \end{center}
\end{figure}

For the heavy Higgs production at LHC, the dominant channel is the gluon fusion process via the top partner loop. The Yukawa coupling involved is approximately $Y_t \sin\alpha_R^t$; as stated above, this right-handed fermion mixing angle is generally very large, of order one and therefore this production process is not suppressed  whereas the top loop is relatively suppressed by the scalar mixing angle $\varepsilon$ or left-handed fermion mixing angle $\alpha_L^t$, as shown in Eq.~(\ref{topyukawa})). The scatter plot of the production cross section is depicted in Fig.~\ref{Hproduction}. For a heavy Higgs with mass of 1 TeV, with 100 fb$^{-1}$ of 14 TeV data, we can expect thousands of heavy Higgs to be produced at LHC. For heavier $H$, the cross section drops rapidly.

We also wish to note that if $M_H > 2M_F$, new decay modes open up. However, for a large range of parameters of the model, the mass of heavy Higgs boson is not large enough to produce heavy top partner pairs. On the other hand, the heavy bottom and tau partners, which are lighter than heavy top partner, could in principle be produced but these channels are suppressed by the small scalar or light-heavy fermion mixings. Therefore, the heavy fermion pair channels are always suppressed, with the branching ratio generally of order $10^{-3}$. We also note that, the $Z-Z'$ mixing effects are suppressed by
$M_Z^2 / M_{Z'}^2$
and are therefore very small. We ignore these effects here.

\section{Neutrinos}
In this section, we briefly address the scale of neutrino masses in the universal seesaw models.
There are two options for neutrino masses in general universal seesaw models: one way is not to introduce a vector-like gauge singlet field ${\cal N}$ but rather obtain neutrino masses as a loop effect~\cite{CM, babu}; the second option is to introduce the vector-like gauge singlet field ${\cal N}_{L,R}$. Let us discuss both options below.

\subsection{ Loop induced neutrino mass} In the case of the first option,
there are two loop diagrams connects the left and right handed chirality neutrino states and lead to a Dirac mass for the neutrinos. It can be estimated to be:
\begin{eqnarray}
m_{\nu_a}\simeq \frac{g^4_2}{(16\pi^2)^2}\frac{m_tm_bm_{\ell_a}}{M^2_{W_R}}I(M_{P_3}, M_{N_3},M_{W_R}) \,.
\end{eqnarray}
This formula gives neutrino masses of order $\sim 0.1$ eV for the $\nu_\tau$ and a factor of 10 smaller for the muon neutrino mass, which are in of the right order of magnitude. Here the expression $I$ is of order $\sim 1$ for our choice of parameters. The neutrino mass hierarchy in this case is normal.
We do not discuss the lepton mixing patterns for this case in this paper.

\subsection{Adding a singlet ${\cal N}_{L,R}$} The second option is to include three gauge singlet fermions  ${\cal N}_a$ and allow either (a) only Dirac mass $M_{\cal N}$ or (b) both Dirac mass $M_{\cal N}$ and
Majorana masses $M_{L,R}$  for ${\cal N}$ fields. Note that both type of mass terms for the ${\cal N}$ field are gauge invariant since $ {\cal N}$ is a gauge singlet field.

If we have only Dirac masses $M_N$ for the vector-like neutral fermions (in a manner similar to the charged fermions), small neutrino masses will require very large $M_N\sim 10^{10}$ GeV for $Y_\nu\sim 10^{-2}$ or very small Yukawa couplings ($Y_\nu \sim 10^{-7}$) for $M_N\sim v_R$. The theory will then have exact lepton number conservation. We do not address the question of neutrino mixings since by arranging an appropriate flavor structure for the neutrino Yukawa couplings $Y_\nu$'s, one can always get the desired pattern.

Coming to the case with both Dirac and Majorana masses for the ${\cal N}$, the neutrino mass matrix in this case reads, in the basis of ($\nu$, ${\cal N}$, $\nu^C$, ${\cal N}^C$) (where all fields are left handed),
\begin{eqnarray}
\left( \begin{matrix}
0 & 0 & 0 & \frac{1}{\sqrt2} Y v_L \\
0 & M_L &  \frac{1}{\sqrt2} Y^T v_R & M_N \\
0 & \frac{1}{\sqrt2} Y v_R & 0 & 0 \\
\frac{1}{\sqrt2} Y^T v_L & M_N & 0  & M_R
\end{matrix} \right) \,.
\end{eqnarray}
In the parameter regime where $M_R \sim M_L \gg M_N \gg Yv_R \gg Yv_L$, the light left handed neutrino masses are given by ${\cal M}_\nu \sim -\frac12 v^2_L Y M^{-1}_R Y^T$. For $M_N\leq M_{L,R}$, the formulae are roughly
\begin{eqnarray}
{\cal M}_{\nu}\simeq - \frac12 v^2_LY\left(M_R-M^T_NM^{-1}_LM_N\right)^{-1}Y^T
\end{eqnarray}
and for right handed neutrinos ($\nu^c$'s), replace $L \leftrightarrow R$ in the above formulae. Naively one might think that in the Majorana alternative, the right handed neutrino masses will be $(v^2_R/v^2_L)$ times those of the left-handed neutrinos (roughly 100 times larger). However this is true only if parity symmetry is exact. If we take the the Majorana mass terms for ${\cal N},\,{\cal N}^c$  to be different and therefore break parity softly, they could have very different forms i.e.mass  scales as well as textures. Therefore by adjusting these terms, one can make the right handed $(\nu^c)$ mass terms in the 10-100 GeV range, and keep them in conformity with cosmology and low energy weak constraints. As an example, consider the case where the magnitudes of all elements of  ${ M}_R$ are in the range of $10^{10}$ GeV and those of $M_L$ are in the TeV range, in this case, the light ``left-handed'' neutrinos  can have sub-eV masses as observed with right handed neutrino masses being in the 100 GeV range.
Since the neutrinos in this case are Majorana fermions, they would give rise to neutrino-less double beta decay.  In this case the collider phenomenology is similar to the conventional TeV scale left-right seesaw case~\cite{TeVLR}. Our goal in this paper is simply to demonstrate that getting small neutrino masses does not pose any challenge to the viability of these models. We postpone further discussion of neutrino masses and accompanying phenomenology to a future publication. We also do not explore the question of whether three singlet fermions ${\cal N}$ are essential, since this is beyond the scope of the paper.

We also wish to comment that we have checked the effect of the neutrino couplings on the vacuum stability question. For $Y_\nu \sim 10^{-2}$ that we assume the running effects are small and in fact their contribution is even less for higher values for $M_{L,R}$ since their contributions to RGEs start only at high scales.

\section{Summary}
We have discussed the question of vacuum stability in the left-right seesaw embedding of standard model with universal seesaw implemented by TeV scale vector-like fermions. This model has only one extra scalar Higgs coupling beyond the standard model and it helps to stabilize the electroweak vacuum till GUT scale. This model has only two neutral Higgs bosons. Identifying the lighter of them with the 126 Higgs boson of standard model, the heavy Higgs mass is found to be below the $v_R$ scale. For parity breaking scale in the few TeV range, it can be accessible at the LHC. We discuss its collider phenomenology such as production cross section and decay properties. We also find  that the vector-like top, bottom and $\tau$ partners $P_3, N_3, E_3$ are below $v_R$ making them LHC accessible. We also find an interesting relation between the three heavy Higgs boson decay modes: $H\to hh, WW, ZZ$, which can provide a test of this model once the heavy Higgs boson is discovered.

\section*{Acknowledgement}
The work of R. N. M. is supported by the National Science Foundation grant No. PHY-1315155. This work of Y. Z. is supported in part by the National Natural Science Foundation of China (NSFC) under Grant No. 11105004.

\end{document}